\documentclass[aps,twocolumn,prb,showpacs,floatfix,amsmath,amssymb,superscriptaddress]{revtex4-1}
\usepackage{graphicx}
\usepackage{bm}
\usepackage{hyperref}
\usepackage{color}
\usepackage{natbib}
\usepackage{changes,cancel,amsmath, amssymb}
\usepackage{braket,float,soul}
\usepackage{comment}

\graphicspath{{./Figs/} } 
\begin{document}
\title{Coherent charge and spin oscillations induced by local quenches\\ in nanowires with spin-orbit coupling}

\author{F. Cavaliere}
\affiliation{Dipartimento di Fisica, Universit\`a di Genova, 16146 Genova, Italy}
\affiliation{SPIN-CNR, 16146 Genova, Italy}

\author{N. Traverso Ziani}
\affiliation{Dipartimento di Fisica, Universit\`a di Genova, 16146 Genova, Italy}
\affiliation{SPIN-CNR, 16146 Genova, Italy}

\author{F. Dolcini}
\affiliation{Dipartimento di Scienza Applicata e Tecnologia, Politecnico di Torino, 10129 Torino, Italy}
\author{{M. Sassetti}}

\affiliation{Dipartimento di Fisica, Universit\`a di Genova, 16146 Genova, Italy}
\affiliation{SPIN-CNR, 16146 Genova, Italy}

\author{{F. Rossi}}
\affiliation{Dipartimento di Scienza Applicata e Tecnologia, Politecnico di Torino, 10129 Torino, Italy}

\begin{abstract}
When a local and attractive   potential is quenched in a nanowire, the spectrum changes its topology from a  purely continuum to  a continuum and discrete portion.
We show that, under appropriate conditions, this quench leads to stable coherent oscillations in the   observables time evolution. In particular, we demonstrate that ballistic nanowires with spin-orbit coupling (SOC) exposed to a uniform magnetic field are especially suitable to observe this effect. Indeed, while in ordinary nanowires the effect occurs only if the strength $U_0$ of the attractive potential is sufficiently strong, even a weak value of $U_0$ is sufficient in SOC nanowires. Furthermore, in these systems coherent oscillations in the spin sector can be generated and controlled electrically by  quenching the gate voltage acting on the charge sector.
We interpret the origin of this phenomenon, analyze the effect of variation of the chemical potential and the switching time of the quenched attractive potential, and address possible implementation schemes.
\end{abstract}

\maketitle
\section{Introduction}\label{sec:intro}
\noindent The race for quantum technologies requires the real-time manipulation of quantum systems to be performed in a controlled way and as rapidly as possible. The analysis of out-of-equilibrium  dynamics thus plays a crucial role in  view of the so-called quantum supremacy.  In particular, when a quantum system is sufficiently well isolated from {dissipative baths} and one or more of its parameters are varied over time,~\cite{qquench1,qquench2,eisert_natphys_2015}  a quantum quench is performed. Quantum quenches can be either global or local. While in the former case the quench parameters extend over the whole system so that an extensive amount of energy is injected into the system,~\cite{gge3,mbl5,gq1,gq2,gq3,gge8} in the latter case the quench is localized in a limited portion of the system, such as  single sites in  lattice models or resonant levels coupled to one-dimensional systems.~\cite{loc1,loc2,loc3,loc4,loc5,loc6,loc7,loc8,LC} Lately, the analysis of local quenches has attracted much interest in view of the long-time crossover between different phases.~\cite{loc4,loc8} 
 
When a quantum quench is performed, the crucial question is whether the isolated system  eventually reaches a steady state and, if so, which one and how. The eigenstate thermalization hypothesis (ETH)~\cite{rigol_nature_2008} states that a generic isolated quantum system  does relax to a steady state that is independent of the initial state and is well described by  standard statistical mechanics. There are, however, relevant cases where the time evolution deviates from such behavior.
In integrable systems~\cite{integra}, for instance, the steady limit of local observables is described in terms of a generalized Gibbs ensemble (GGE)~\cite{gge1,gge2,gge3,gge4,gge5,gge6,gge7,gge8,gge9,gge10}  that can be constructed out of the local conserved quantities of the system. This enables one  to predict the long-time behavior of quenched integrable systems even in the presence of non-equilibrium quantum phase transitions.~\cite{qpt1,qpt2} 
Another relevant situation where ETH does not hold is in interacting  disordered systems, due to  the  mechanism of many-body localization:~\cite{mbl1,mbl2,mbl3,mbl5} energy exchange among localized states becomes suppressed and thus the system retains local memory of the pre-quench state. The out of equilibrium dynamics after a quench is thus no trivial question to answer.\\

There exist two main  platforms where quantum quenches can be implemented and investigated. Cold atom systems are considered  an excellent example, as they can be effectively decoupled from the environment and their parameters can be controlled in real time  with high accuracy, typically by means of lasers~\cite{ca1,ca2,ca3,ca4,ca5}. However, condensed matter systems are more straightforwardly integrated with conventional electronics, and offer  the possibility to analyze transport properties in a quite natural way and to realize a quench electrically, e.g. by simply varying the gate voltage of a nearby metallic gate electrode. Moreover, condensed matter systems characterized by strong spin-orbit coupling (SOC) are also ideal candidates for spintronics~\cite{awschalom1,streda,awschalom2,soc2,soc5}, as information encoded in the spin degrees of freedom can be manipulated electrically.

In this context,  Rashba SOC nanowires (NWs), such as  InSb~\cite{nilsson_2009,deng_2012,kouwenhoven_PRL_2012,weperen} or InAs~\cite{ensslin_2010, gao_2012,joyce_2013,scherubl,soc4} NWs, have recently attracted a lot of interest, due to their peculiar properties. When a magnetic field is applied along the NW axis, the electronic states inside the thereby created magnetic gap mimic quite well the helical edge states of quantum Spin-Hall effect~\cite{kane-mele2005a,kane-mele2005b,bernevig_science_2006}, as electrons propagating in opposite directions along the NW also have opposite spin orientations. Furthermore, when a superconducting film is deposited on top of a SOC NW, the two ends of the NW exhibit the appearance of Majorana  modes, i.e. quasi-particles that are equal to their anti-particles and that have non-trivial braiding properties, which    make them ideal candidates for quantum computational purposes~\cite{vonoppen_2010,dassarma_2010,alicea_2012,kouwenhoven_2012,liu_2012,heiblum_2012,xu_2012,defranceschi_2014,marcus_2016,marcus_science_2016}. 
So far most  of these investigations have focussed on equilibrium properties~\cite{ojanen_2012, Dolcini-Rossi_PRB_2018} or steady state transport.\cite{rainis-loss_PRL_2014, rainis-loss_PRB_2014,aguado_2015,soc4}
Concerning time-dependent perturbations, while the effects of periodic driving have been discussed~\cite{loss_PRL_2016},  the case of quantum quenches in these systems is much less explored. It is known, for instance, that when the magnetic field is suddenly applied the magnetization exhibits a non-monotonic response.~\cite{freeze}
 
In this paper we  study the quench of a {\em local} and {\em attractive} potential in a single-channel ballistic NW in the mesoscopic regime. Our motivation   is that   this   quench changes the  intrinsic structure of the spectrum,  which is purely a continuum when the potential is absent, while   after the quench it also exhibits some discrete energy levels related to the localized bound states. Since the pre-quench state is not an eigenstate of the post-quench Hamiltonian, 
the time evolution of the system crucially depends on the interplay between and within discrete and continuum states. In particular, we show that under appropriate circumstances the system does not reach a steady state. Instead, the NW observables  display long-living coherent oscillations,  which persist after the transient dynamics represented by the propagation of a ``light cone" from the potential.~\cite{LC} The period of these oscillations is related to the 
energy difference between discrete states. Importantly, the presence of the discrete energy levels is a necessary, but not sufficient condition to observe this effect. Indeed, selection rules forbid any transition between consecutive discrete levels of a one-dimensional system. In an ordinary NW with parabolic spectrum, this constraint implies that the strength of the quench attractive potential  must be sufficiently strong to give rise to  a few discrete levels (typically at least three). 

\begin{figure}
	\centering
	\includegraphics[width=0.9\linewidth]{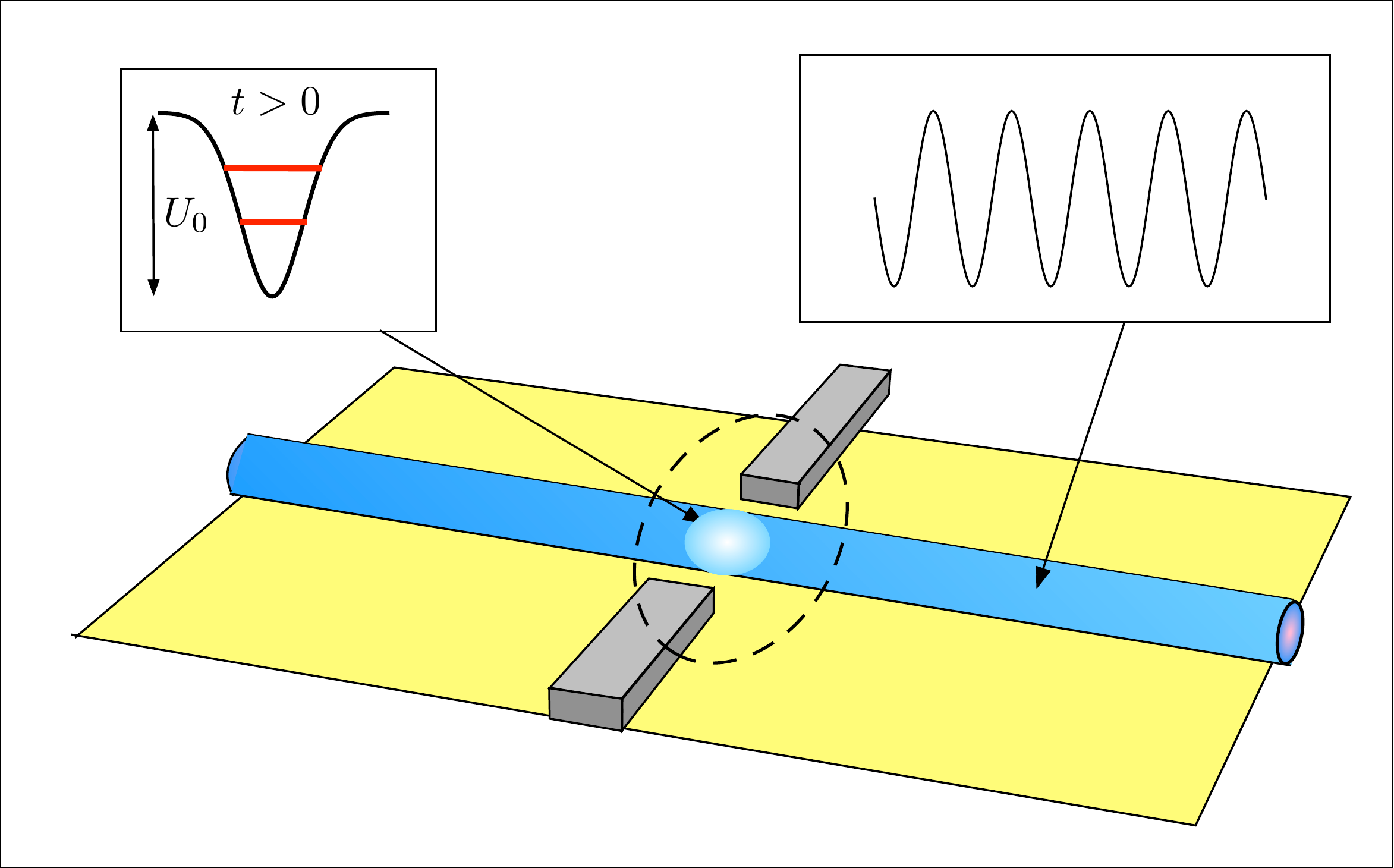}
	\caption{(Color online) Sketch of the setup: a SOC nanowire coupled to narrow metallic gates, which generate  a localized   potential. An attractive gate potential is  quenched at the time $t=0$, generating localized bound states near the gated region, and causing coherent oscillations both in the charge and in the spin sector of the nanowire observables. }
	\label{fig:Fig-setup}
\end{figure}

A much more interesting scenario emerges in SOC NWs. Indeed the interplay between spin and charge degrees of freedom has two important implications. First, the threshold of the potential strength   for the coherent oscillations to emerge is strongly reduced. Second, in this system a quench on the charge degree of freedom, performed e.g. simply by a gate voltage switch in a nearby finger gate [as sketched in Fig.\ref{fig:Fig-setup}], directly generates persistent oscillations in the spin channel. We describe this phenomenon in details and also discuss  the space evolution of these oscillations along the NW. Furthermore, we investigate its robustness to variation of the chemical potential and the switching time of the attractive potential.

The paper is organized as follows: in Sec.~\ref{sec:simplemodel} we illustrate the idea underlying the emergence of long-living oscillations caused by the    quenched attractive potential, pointing out the limitations of the case of a conventional NW. In  Sec.~\ref{sec:socwire} we introduce the model for  SOC NWs and describe the method we use to investigate the dynamics upon the quench. Then, in  Sect.~\ref{sec:results}, we present our results  and show  that SOC are ideal to observe the effect. We discuss the time evolution of charge and spin sector observables under the quench of a localized attractive Gaussian potential, interpret the results and address the effects of variation of the chemical potential and a finite switching time in the quench protocol. Finally, conclusions are drawn in Sec.~\ref{sec:conclusions}.

\section{The emergence of quench-induced oscillations}\label{sec:simplemodel}
To illustrate the mechanism underlying the emergence of coherent and long-living oscillations, we consider a NW directed along the $x$ axis. Within the customary effective mass $m^*$ approximation, the NW  parabolic dispersion relation is $\varepsilon(k)=\hbar^2 k^2/2 m^*$, and the spectrum before the quench   consists of a continuum of propagating plane waves labelled by the wavevector~$k$. 
Suppose e.g. that the system is initially prepared in the many-body ground state of the pre-quench Hamiltonian, filling up all levels up to a given chemical potential value $\mu$.
At the time $t=0$ an attractive potential~$U$, localized over a lengthscale $\lambda_U$   around the origin $x=0$, and characterized by a potential strength $U_0$, is switched on. 
As a definite example one can consider the sudden quench of an attractive Gaussian profile {$U(x,t)=-\theta(t)\, U_0\, \exp[-x^2/2\lambda_U^2]$}, where~$\theta$ indicates the Heaviside function. In the presence of such localized potential, the system loses the translational invariance, and the wavevector $k$ is no longer a good quantum number.  In addition to the continuum part of the spectrum,  some discrete energy levels  $\varepsilon_{1}<\ldots<\varepsilon_{N}$ appear after the quench, whose number~$N$ depends on the values  of $U_0$ and $\lambda_U$. If the attractive potential  is spatially even, $U(x,t)=U(-x,t)$, like the Gaussian one,  the discrete energy levels $\varepsilon_n$ have an alternating spatial parity {$p_n=(-1)^{n+1}$ ($n=1,\ldots,N$)} the ground state being even ($p_1=+1$).  Furthermore, in an ordinary NW, each level has a trivial  two-fold spin  degeneracy $s=\pm$.\\

The post-quench electron field operator with spin~$s$ thus consists of a superposition of  the continuum (c)   states $\phi_{c,s}(x)$ {-- with $c$ a continuum index --} and the $N$ discrete  localized states $\chi_{n,s}(x)$, both equipped with the related fermionic creation operators~$\gamma^\dagger$,
\begin{equation}\label{eq:fermiopsim}
\psi^{\dagger}_s(x)= \int{\mathrm{d}} c \, \phi_{c,s}(x) \gamma^{\dagger}_{c,s}\,+ \sum_{n}\chi_{n,s}(x)\gamma^{\dagger}_{n,s} \quad.
\end{equation}
The time evolution of any NW observable depends on all combinations of these two types of components. Let us consider, for instance,  the time evolution of the  charge density expectation value, given by
\begin{equation}
\rho(x,t)=\langle\Psi^{\dagger}(x,t)\Psi(x,t)\rangle_0 =\sum_s\langle\psi^{\dagger}_s(x,t)\psi^{}_s(x,t)\rangle_0   \,,	
\end{equation}
where   $\langle\ldots\rangle_0$ denotes the quantum average on the pre-quench   ground state of $H(t<0)$, while the time evolution is governed by $H(t>0)$. Since the NW Hamiltonian is diagonal in spin, the two spin sectors simply decouple and the density consists of three terms 
\begin{equation}\label{eq:rhosm}
\rho(x,t)=\sum_{s}\left[\rho_s^{(\mathrm{cc})}(x,t)+\rho_s^{(\mathrm{dc})}(x,t)+\rho_s^{(\mathrm{dd})}(x,t)\right]\,,	
\end{equation}
where
\begin{equation}\label{eq:rhoccsm}
\rho_s^{(\mathrm{cc})}(x,t)\!=\!\!\iint{\mathrm d}c\,{\mathrm d}c' \, \phi^{*}_{c,s}(x)\phi_{c',s}(x) \,e^{\frac{i}{\hbar}[\varepsilon(c')-\varepsilon(c)]t}C_s(c,c')\nonumber
\end{equation}
\begin{equation}\label{eq:rhodcsm}
\rho_s^{(\mathrm{dc})}(x,t)=\sum_{n}\!\int{\mathrm d}c \, \chi^{*}_{n,s}(x)   \phi_{c,s}(x) \, e^{\frac{i}{\hbar}[\varepsilon_{n}-\varepsilon(c)]t}B_s(n,c')\nonumber
\end{equation}
\begin{equation}\label{eq:rhoddsm}
\rho_s^{(\mathrm{dd})}(x,t)=\sum_{n,n'}\chi^{*}_{n,s}(x)\chi_{n',s}(x)\, e^{\frac{i}{\hbar}[\varepsilon_{n}-\varepsilon_{n^\prime}]t}A_s(n,n')\nonumber
\end{equation}
and
\begin{eqnarray}
A_s(n,n')&=&\langle \gamma^{\dagger}_{n,s} \gamma^{}_{n',s}\rangle_0\,\label{eq:Asm} \\
B_s(n,c)&=&\langle \gamma^{\dagger}_{n,s} \gamma^{}_{c,s}\rangle_0\,,\label{eq:Bsm} \\
C_s(c,c')&=&\langle \gamma^{\dagger}_{c,s} \gamma^{}_{c',s}\rangle_0\,.\label{eq:Csm}
\end{eqnarray}

\noindent Since the ground state of the pre-quench Hamiltonian $H(t<0)$ is not an eigenstate of the post-quench Hamiltonian $H(t>0)$, the expectation values of the post-quench bilinears over the pre-quench ground state, Eqs.(\ref{eq:Asm}), (\ref{eq:Bsm}) and (\ref{eq:Csm}), are in general non vanishing, 
 and so are $\rho_s^{(\mathrm{dc})}(x,t)$ and $\rho_s^{(\mathrm{cc})}(x,t)$ and $\rho_s^{(\mathrm{dd})}(x,t)$. In particular, the term $\rho_s^{(\mathrm{cc})}(x,t)$ describes the usual ``light cone effect" observed in quenches:~\cite{LC} two counter-propagating dispersive density bumps originating from $x=0$  {travel}  as a wave, transferring the quench information  along  the NW. Because this term contains an infinite number of  phase factors oscillating in time with a continuum of frequencies $[\varepsilon(c')-\varepsilon(c)]/\hbar$, in the long time behavior a dephasing of all $c \neq c^\prime$ components occurs, and $\rho_s^{(cc)}(x,t)$ saturates to the value given by  the diagonal terms $c=c^\prime$ only,  $\rho_s^{(\mathrm{cc})}(x,t)\to\tilde{\rho}_s^{(\mathrm{cc})}(x)$.\\

The term $\rho_s^{(\mathrm{dc})}(x,t)$ represents the mixed discrete-continuum contribution. Similarly to the previous term, it also contains a continuum of frequencies, depending on the energy separation between each discrete level  with the whole set of continuum states, which dephase causing 
 damped oscillations. In the long  time limit this term eventually vanishes.~\cite{powerlaws}
Finally, the term $\rho_s^{(\mathrm{dd})}(x,t)$, related to the discrete states  localized near the attractive potential,  can be rewritten as $\rho_s^{(\mathrm{dd})}(x,t)=\tilde{\rho}_s^{(\mathrm{dd})}(x)+\delta\rho_s^{(\mathrm{dd})}(x,t)$. While the first term, defined as $\tilde{\rho}_s^{(\mathrm{dd})}(x)=\sum_{n}A_s(n,n)\left|\chi_{n,s}(x)\right|^2$, and stemming from the diagonal  contribution ($n=n^\prime$), is time-independent and  yields a steady term to the density, the second term, defined as
\begin{equation}\label{eq:osc}
\delta\rho_s^{(\mathrm{dd})}(x,t)=\sum_{n\neq n'} \chi^{*}_{n,x}(x)A_s(n,n')\chi_{n',s}(x)e^{i[\varepsilon_{n}-\varepsilon_{n'}]t/\hbar}\,, 
\end{equation}
oscillates in time. In sharp contrast to the previous time-dependent terms discussed above, which involve a continuum set of frequencies and eventually dephase in the long time limit, the term $\delta\rho_s^{(dc)}(x,t)$ only involves a few frequencies  $(\varepsilon_{n}-\varepsilon_{n'})/\hbar$ related to the finite energy differences between the discrete bound states. As a consequence, such term describes a localized density perturbation that coherently oscillates without any dephasing and   survives after the ``light cone" has propagated away from the attractive potential. 

From the very structure of the term in Eq.(\ref{eq:osc}), one can see that a necessary condition for the effect to arise is that the potential strength $U_0$ must be strong enough to give rise to at least $N=2$ non-degenerate levels. There are, however,   selection rules  that further increase the actual minimal number of levels. 
Indeed in a realistic case where the attractive potential is created by a finger gate voltage, it is reasonable to assume that such potential is well described by a spatially even  attractive potential $U(x,t)=U(-x,t)$, like the Gaussian one mentioned above. Then, since any two {\it consecutive} discrete states have opposite parity, they cannot be coupled to each other  and the effect vanishes. This can also be seen {in a more hand-waving way} e.g.  at the origin $x=0$ by realizing that, if only two levels are present, Eq.(\ref{eq:osc}) vanishes, since the second bound state wavefunction $\chi_{2,s}$ has a node at $x=0$.  Therefore, in order for the coherent oscillations to emerge from Eq.(\ref{eq:osc}), one needs $A_s(n,n')\neq 0$ with $n\neq n'$ and equal parity ($p_n p_{n^\prime}=+1$). Since the parity of bound states $\varepsilon_n$ alternates with $n$, the actual realistic minimal required number of discrete levels  is $N=3$. Thus, there is a threshold on the impurity strength $U_0$ to induce this effect.

\begin{figure}
	\centering
	\includegraphics[width=1\linewidth]{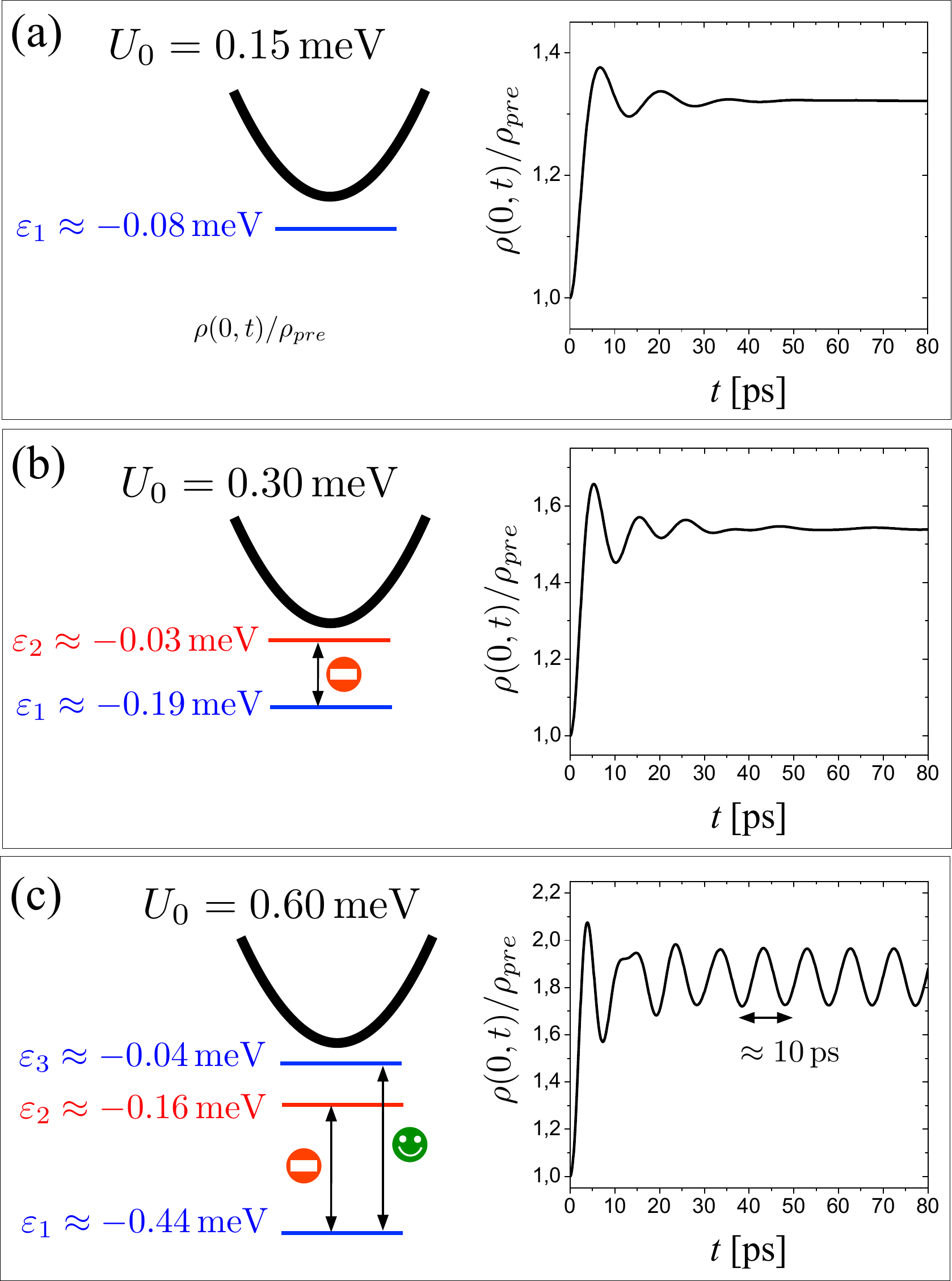}
	\caption{(Color online) The case of a conventional NW after the quench of   an attractive Gaussian potential with $\lambda_U=150\,{\rm nm}${, $\mu=0.2\,{\rm meV}$} and three different values of the potential strength: (a) $U_0=0.15 \,{\rm meV}$, (b) $U_0=0.30 \,{\rm meV}$ and (c) $U_0=0.60 \,{\rm meV}$. The left panels shows the corresponding number of localized bound states, the color denoting the parity (blue for $p=+1$ and red for $p=-1$). The parabolic band of the pre-quench Hamiltonian is shown as a guide to the eye. The right panels display the corresponding behavior of the electron density at the origin (center of the quenched potential) {normalized to its pre-quench value $\rho_{pre}$,} the left panels depict schematically the energy of the bound states per spin band and the parity $p_n$ (blue for +1, cyan for -1).}
	\label{fig:Fig-ESO=0}
\end{figure}

These arguments are illustrated in Fig.\ref{fig:Fig-ESO=0} for the case of a quenched Gaussian attractive potential centered around the origin over a lengthscale $\lambda_U=150\,{\rm nm}$. The three panels show the post-quench time evolution of the density $\rho(0,t)$ at the origin $x=0$, for three different values of the Gaussian potential strength $U_0${, normalized to the constant pre-quench density $\rho_{pre}=\rho(x,t<0)$}. Panel (a) displays the case of a weak potential $U_0=0.15 \,{\rm meV}$, generating only one bound state. As can be seen, the density exhibits strongly damped oscillations in the transient and, in the long time limit, the density approaches a steady state.  Even when the attractive potential strength is increased to the value $U_0=0.30 \,{\rm meV}$, where two bound states are present  [see panel (b)], the same qualitative behavior occurs, as the two states are decoupled by parity selection rules. However, for the stronger value  $U_0=0.6\,{\rm meV}$ displayed in panel (c),   three bound states appear, and the density exhibits   stable oscillations, with a period $\tau=2\pi \hbar/\Delta\varepsilon\approx 10\,{\rm ps}$ which corresponds to the inverse energy separation $\Delta\varepsilon=0.4\ \mathrm{meV}$ between the first and the third bound states sharing the same parity $p=+1$.\\

To summarize, the  appearance of the density oscillations discussed here is due to i) the fact that the pre-quench state is a complex excitation of the post-quench Hamiltonian, and ii) the presence of various discrete energy levels corresponding to bound states, which effectively have an ``internal dynamics" characterized by a few frequencies and decoupled from the continuum states, which instead exhibit a long time dephasing.\\

\subsection{Effects of a uniform magnetic field} 
As discussed above, the coherent oscillation effect arises only if the strength $U_0$ of the attractive potential overcomes the threshold needed to generate at least $N=3$ localized levels. One may naively think of effectively reducing this threshold by applying a uniform magnetic field, which would lift the spin degeneracy ($\varepsilon_n \rightarrow \varepsilon_{n,\uparrow}, \varepsilon_{n,\downarrow}$) and thereby double the number of the localized states. However, because the quench potential purely acts in the {\em charge} channel,  in a conventional NW lacking of any internal spin-structure each spin species is dynamically  decoupled from the other. Thus, if the coherent oscillation effect is not present in the absence of a magnetic field, it cannot be induced by its application. As an illustrative example, if the magnetic field is applied e.g. along the NW axis and the chemical potential is set to $\mu=0$, so that the higher Zeeman split band is empty before quenching the attractive potential, it will remain so even after the quench, and the polarization will be constantly equal to 1 at any time. More in general, in conventional NWs exposed to a uniform magnetic field the spin polarization components that are orthogonal to the magnetic field direction will always remain vanishing even after the quench. In the following sections we shall show that the situation is quite different in a SOC NW, where the interplay between SOC and a uniform magnetic field generates an effectively  inhomogeneous magnetic field that couples the states. \\

\section{Spin-orbit coupled nanowires}\label{sec:socwire}
While in the previous section we have outlined the general idea underlying the emergence of long-living coherent oscillations, in this section we show that SOC NWs are ideal to observe such effect.
\subsection{The model}\label{sec:genremarks}
Spin-orbit coupled NWs, such as InSb and InAs, are characterized by a strong spin-orbit coupling, mainly arising from structural inversion asymmetry due to the   underlying substrate, which leads the electron flowing along the NW to experience an effective `magnetic field'.  The NW Hamiltonian can be written as $\hat{\mathcal{H}}=\int dx \, \Psi^\dagger(x) H \,  \Psi^{}(x)$, where $\hat{\Psi}(x)= \left(   \psi_\uparrow(x) \,,\,\psi_\downarrow(x) \right)^T$ denotes the electron spinor field, and $\uparrow,\downarrow$ correspond to spin projections along positive and negative spin-orbit effective `magnetic' field direction. Here below we shall describe the pre- and the post-quench single-particle Hamiltonian operator $H$.\\

{\it Pre-quench Hamiltonian.} Before switching the  attractive potential the NW  Hamiltonian is 
\begin{equation}  \label{H0}
H_0= \left( \frac{p_x^2}{2 m^*} -\mu\right) \sigma_0 -\frac{\alpha}{\hbar} \,   p_x   \sigma_z\, \, - h_x \sigma_x  \quad,
\end{equation}
where  $p_x=-i \hbar \partial_x$ is the momentum operator, $\mu$ the chemical potential, $\alpha$ the coupling constant of the Rashba spin-orbit 'effective magnetic field' and $z$ its direction, while $h_x>0$ is the Zeeman energy related to an actual  external magnetic field directed along the NW axis~$x$. Finally, $\boldsymbol{\sigma}=(\sigma_x, \sigma_y,\sigma_z)$ are Pauli matrices and $\sigma_0$ the $2 \times 2$ identity matrix. 

The spin-orbit term and the magnetic field lead to a splitting of the otherwise degenerate bands of an ordinary NW, so that the spectrum of $H_0$ consists of two bands $\nu=\pm$
\begin{equation}\label{pre-quench-eigval}
\varepsilon_{\pm}(k)=\frac{\hbar^2 k^2}{2m^*}\pm\sqrt{h_x^2+\alpha^2 k^2}	
\end{equation}
In particular, while the upper  band ($\nu=+$) always has only one minimum $\varepsilon^{\min}_{+}=h_x$ at $k=0$, the structure of the lower band  ($\nu=-$) strongly depends on the relative weight of the Zeeman energy $h_x$ and the   spin-orbit energy  
\begin{equation}
\label{ESO-def}
E_{SO}=\frac{m^{*}\alpha^2}{2\hbar^2}\quad.\end{equation}
Explicitly, one finds \cite{Dolcini-Rossi_PRB_2018} 
\begin{equation}
\varepsilon^{\min}_{-}=\left\{ \begin{array}{lll} 
-h_x & \mbox{at } k=0 & \mbox{if } E_{SO}<\frac{h_x}{2} \\ & & \\
 -E_{SO}-\frac{h_x^2}{4 E_{SO}} & \mbox{at } k=\pm k^{min} & \mbox{if } \frac{h_x}{2} < E_{SO} 
\end{array}
\right.
\end{equation}
i.e., while for strong magnetic field there is only one minimum at $k=0$, for weaker magnetic field two degenerate minima are present   at $k=\pm k^{min}$ where $k^{min}= k_{SO} (1-h_x^2/4 E_{SO}^2)^{1/2}$ and $k_{SO}=m^* |\alpha|/\hbar^2$ is the spin-orbit wavevector.
In particular, in the SOC dominated regime ($E_{SO} \gg h_x/2$) one has $\varepsilon^{\min}_{-} \approx -E_{SO}$ and $k^{min}\approx   k_{SO}$. The spin-orbit wavevector $k_{SO}$ also determines {the spin-orbit length scale over which spin precesses
\begin{equation}\label{lSO-def}
l_{SO} = k^{-1}_{SO}= \frac{\hbar^2}{m^* |\alpha|}\,.
\end{equation}
}
 
{\it Post-quench Hamiltonian.} When the attractive potential $U$ localized around the origin is quenched, the Hamiltonian 
\begin{equation}
H=H_0+H^\prime(t)
\end{equation}
acquires the additional term  
\begin{equation}\label{Hprime}
H^\prime(t)=U(x,t)\sigma_0=f(t)\, U(x) \sigma_0\quad,
\end{equation}
where $f(t)$ is a generic dimensionless function, increasing from 0 to 1 and describing the quench protocol. While most results will be given in Sec.\ref{sec:results} for the case of a sudden quench,   $f(t)=\theta(t)$,  in Sec.~\ref{sec:finite-tau} we shall explicitly discuss the effect of a finite switching time in the quench protocol. Concerning the spatial shape $U(x)$ of the attractive potential, we assume like in the previous section that it extends over a lengthscale $\lambda_U$ around the origin, and that it is characterized by a maximal strength $U_0$. 

\subsection{Symmetries}
We observe that various discrete symmetries are broken in the pre-quench Hamiltonian $H_0$: the spin-orbit term breaks both parity $\mathcal{P}$ and $\sigma_x$-spin symmetries,  whereas the magnetic field  breaks time-reversal symmetry. However, it is straightforward to verify that the product $\Lambda=\mathcal{P}\sigma_x$ still commutes with the pre-quench Hamiltonian, $[H_0,\Lambda]=0$. Furthermore,  if the potential is spatially even, $U(x)=U(-x)$, also the post-quench Hamiltonian commutes with the operator  $\Lambda$. In this case a selection rule thus exists: Because this operator has eigenvalues $\lambda=\pm 1$, subspaces characterized by different values of $\lambda$ are completely decoupled even when the local attractive potential is quenched. 

\subsection{Density matrix approach}\label{sec:density-mat}
In order to investigate the dynamical evolution of observables, such as charge and spin densities, we adopted a simulation strategy based on the single-particle density-matrix formalism, successfully employed for the investigation of energy relaxation and decoherence phenomena in semiconductor quantum nanodevices~\cite{dolcini-iotti-rossi_PRB_2013} as well as in carbon-based materials.~\cite{rosati-dolcini-rossi_PRB_2015}
More specifically,  labelling by~$\beta$ the eigenstates of the pre-quench Hamiltonian $H_0$ and by $\varepsilon_\beta$ the corresponding energy levels,
for any single-particle quantity described by the operator $\hat A$ its average value can be written as
\begin{equation}\label{aveval}
A (t)= \sum_{\beta_1\beta_2} \rho_{\beta_1\beta_2}(t) A_{\beta_2\beta_1}\ ,
\end{equation}
where
\begin{equation}\label{spdm}
\rho_{\beta_1\beta_2}(t)= \langle \hat c^\dagger_{\beta_2}(t) \hat c^{ }_{\beta_1}(t) \rangle
\end{equation}
is the usual single-particle density matrix expressed in terms of corresponding creation and annihilation operators acting on the generic eigenstate $\beta$.
The time evolution of the single-particle density matrix in (\ref{spdm}) is obtained  by solving the Liouville-Von Neumann equation
\begin{eqnarray}\label{lvne}
\frac{d \rho_{\beta_1\beta_2}}{d t} 
&=&
\frac{\varepsilon_{\beta_1} - \varepsilon_{\beta_2}}{{i}\hbar}\, \rho_{\beta_1\beta_2}
\nonumber \\
&+&
\left(\frac{1}{{i}\hbar} \sum_{\beta_3}
H^\prime_{\beta_1\beta_3} \rho^{}_{\beta_3\beta_2}
+ \textrm{H.c.}
\right)\ ,
\end{eqnarray}
where $H^\prime_{\beta\beta'}$ denote the matrix entries of the quench Hamiltonian Eq.(\ref{Hprime}) in the pre-quench basis $\beta$.
In the case under investigation,   the pre-quench Hamiltonian is translationally invariant and can be diagonalized exactly. The quantum label $\beta=(k,\nu)$ thus consists of the plane-wave wavevector and the upper/lower band index $\nu=\pm$, the eigenvalues are given in Eq.(\ref{pre-quench-eigval}) and the eigenvectors are\cite{Dolcini-Rossi_PRB_2018}
\begin{eqnarray}
\label{eigenvectors}
w_{k-}= \left(
\begin{array}{c}
\cos \frac{\theta_k}{2} \\  \\ 
 \sin \frac{\theta_k}{2}\,  
\end{array}\right)\hspace{0.5cm} 
w_{k +}= \left(
\begin{array}{c}
-  \sin \frac{\theta_k}{2}\\  \\ 
\cos \frac{\theta_k}{2}
\end{array}\right)\,. \hspace{0.5cm} 
\end{eqnarray}
where the angle $\theta_k$ is given by
\begin{equation}\label{thetak-def}
\left\{\begin{array}{lcl}
\cos \theta_k  &=&\displaystyle  \frac{\alpha k}{\sqrt{(\alpha k)^2+h_x^2}}\\  
\sin \theta_k  &=&\displaystyle \frac{{h_x}}{\sqrt{(\alpha k)^2+h_x^2}}
\end{array}\right.  
\end{equation}
Then, the Liouville-Von Neumann Eq.(\ref{lvne}) is solved  numerically  via a fourth-order Runge-Kutta scheme. To this aim, we adopt a uniform discretization of the wavevector $k$, which   amounts to introducing a periodic-boundary-condition scheme with a characteristic length~$L$; in turn, this determines the existence of  a ``revival time" $\tau_{r}=2L/v_{F}$ for the excitation to turn around the effective `ring', where $v_F$ is the Fermi velocity of the system (in our case   $v_F \sim 10^5 \,{\rm m/s}$). To get rid of these spurious effects, we have taken $L$ much longer than a realistic NW length ($L\approx 16 \,{\rm \mu m}$) and we have performed simulation for time scale shorter than such recursion   time  $\tau_{r}\approx 400\ \mathrm{ps}$. \\

Finally, as a model for the spatial profile for the quench potential  in Eq.(\ref{Hprime}) we {adopt the Gaussian profile} 
\begin{equation}\label{Gauss-profile}
U(x)=-U_0\, e^{-x^2/2\lambda_U^2}
\end{equation}
centered around the origin, extending over a lengthscale $\lambda_U$ and with a typical strength $U_0$ {already considered above}.

\section{Results for SOC nanowires}\label{sec:results}
In this section we present our results about SOC NWs. Our analysis shows that  there are two main reasons why SOC NWs are ideal candidates to observe the effect of long-living coherent oscillations. The first one is that the effect arises already for much lower values of $U_0$ than  in ordinary NWs. The second one is that, despite the quench potential is applied on the charge sector,  the SOC also transfers the effect to the spin sector, so that coherent oscillations in  spin density components arise, which are absent in ordinary NWs.
We explicitly illustrate these features here below in Sec.\ref{eq:ressud} for the case of a sudden quench, while in Sec.\ref{sec:interpretation} we explain why SOC leads to these advantageous features. Then, in Sec.\ref{sec:chem-pot} and Sec.\ref{sec:finite-tau} we discuss the effects of the chemical potential and a ramp with a finite switching time.

\subsection{Numerical values}\label{sub:param}
For definiteness, we shall focus on the case of InSb NWs,~\cite{nilsson_2009,deng_2012,kouwenhoven_PRL_2012,weperen} with $m^{*}=0.015 m_e$  (with $m_e$ the electron mass), and consider a typical spin-orbit energy $E_{SO}=0.25\ \mathrm{meV}$, which corresponds to a $l_{SO}\approx100\ {\rm nm}$.{~\cite{valSO}} We assume that the external magnetic field applied along the NW axis yields a Zeeman energy  $h_{x}=0.2\ \mathrm{meV}$.  Concerning the parameters of the Gaussian profile (\ref{Gauss-profile}) we take a lengthscale $\lambda_U=150\,{\rm nm}$ and a   strength $U_0=0.15\,{\rm meV}$. The pre-quench state is the thermal equilibrium state at low temperature $T=250\ \mathrm{mK}$ and $\mu=0$ (corresponding to the middle of the magnetic gap), unless otherwise specified.

\subsection{Sudden quench}\label{eq:ressud}
We start by discussing the case of a sudden quench, where the quench protocol function of Eq.(\ref{Hprime}) is $f(t)=\theta(t)$.\\

\begin{figure}
	\centering
	\includegraphics[width=1\linewidth]{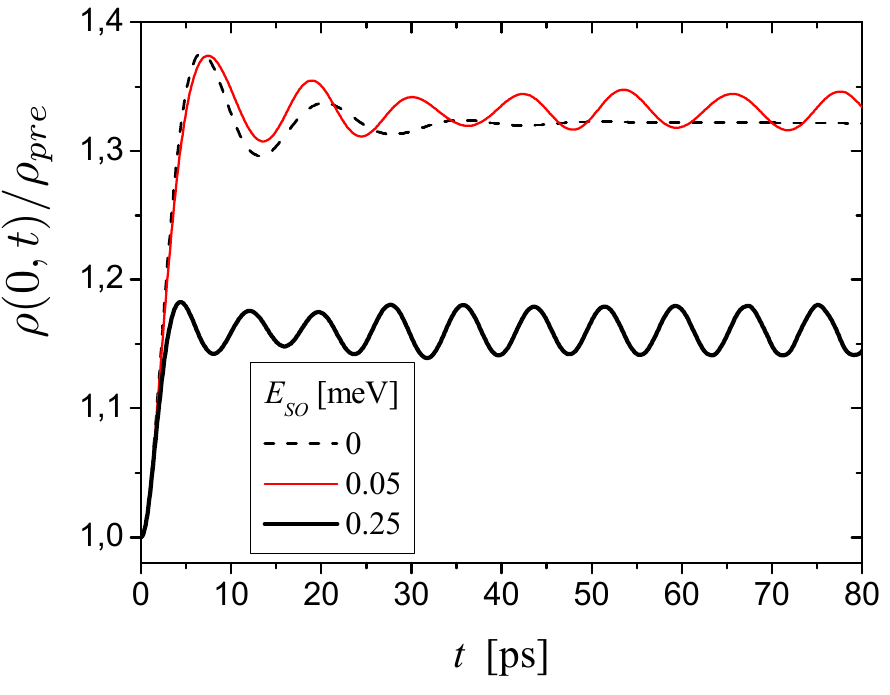}
	\caption{(Color online) The time evolution of the charge density $\rho$ at $x=0$ {(normalized to the constant and uniform pre-quench value $\rho_{pre}$)} after a sudden quench of the attractive Gaussian potential, in the case of a SOC nanowire {with $E_{SO}=0.05\,{\rm meV}$ (thin red solid curve) and $E_{SO}=0.25\,{\rm meV}$ (thick black solid curve),} other parameters given in Sec.\ref{sub:param}: coherent oscillations are clearly visible. For comparison, the case of a conventional nanowire with the same parameters, except for vanishing SOC, is also plotted shown (dashed curve). In that case oscillations rapidly decay.}
	\label{fig:rho}
\end{figure}

\begin{figure}
	\centering
	\includegraphics[width=0.8\linewidth]{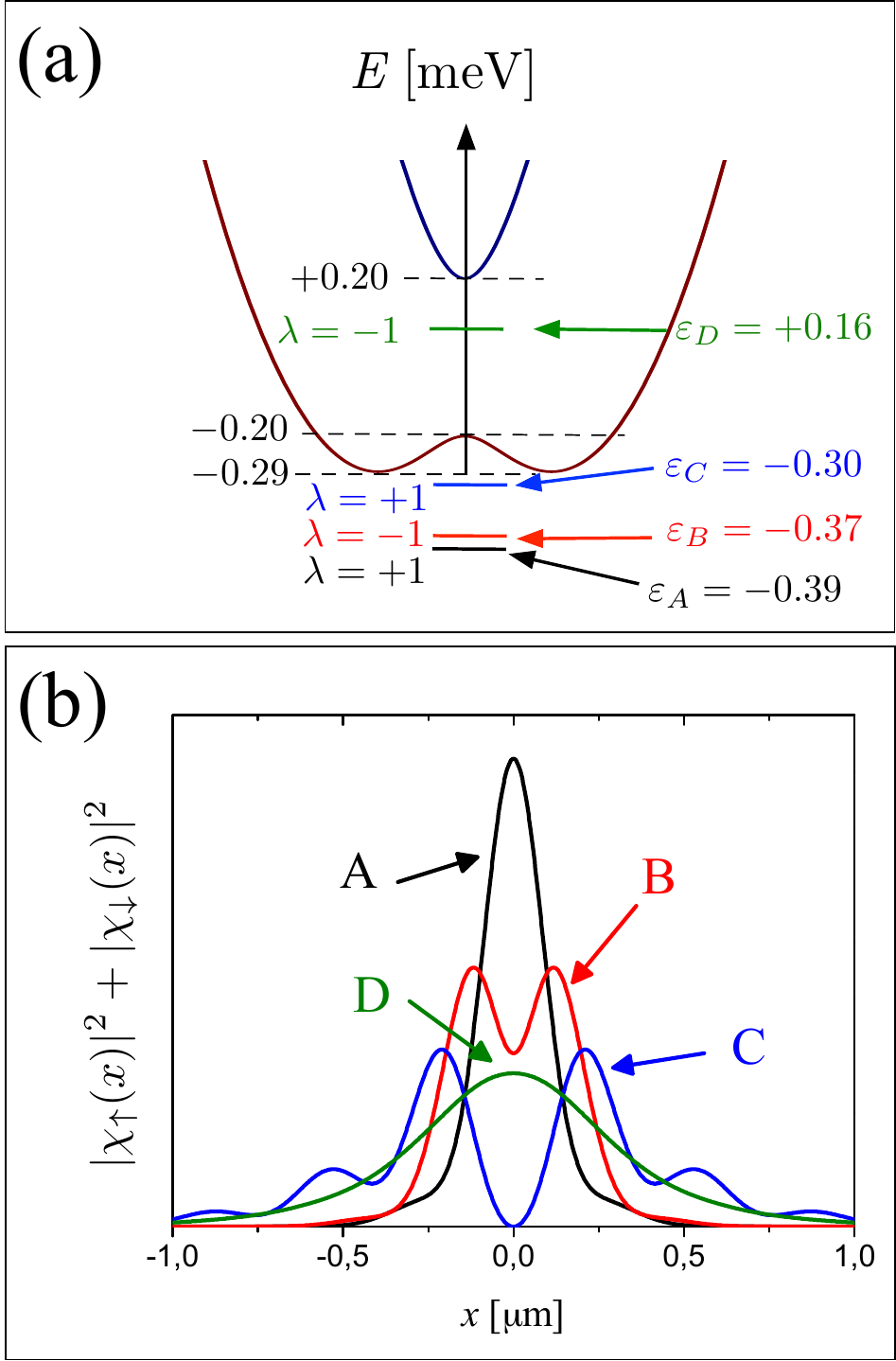}
	\caption{(Color online) (a) Scheme of the four bound states of the post-quench hamiltonian $H'(t>0)$ as found by exact diagonalization. For reference, we plot also the energy bands of the nanowire before the quench. Parameters as in Sec.~\ref{sub:param}. (b) Plot of the density profile of the four bound states.}
	\label{fig:Fig-Wave}
\end{figure}

{\it Charge density.} In Fig.~\ref{fig:rho} the time evolution of the charge density $\rho(0,t)$ at the origin $x=0$ (center of the localized potential) is shown{, normalized to the constant pre-quench value $\rho_{pre}$,} for a SOC NW (solid curves) {and two different values of $E_{SO}$, compared} to the case of a conventional NW (dashed curve) where all other parameters except the SOC are the same. For the value $U_0=0.15\,{\rm meV}$ of the potential strength, a conventional NW only has 1 bound state per spin channel and the oscillations are damped down rapidly due to dephasing, as discussed in Sec.\ref{sec:simplemodel}. 
In contrast, {even for a rather weak value of $E_{SO}=0.05\,{\rm meV}$ and the same value of $U_0$ the SOC NW -- thin red solid curve -- exhibits long-living coherent oscillations. Further increasing $E_{SO}$ -- thick black solid curve -- results in oscillations with larger amplitude. Let us focus on the latter case.} Indeed four bound states are found in the SOC NW, whose energies are schematically depicted in Fig.\ref{fig:Fig-Wave}(a). The NW bands before the quench have been added as a guide to the eye. As one can see, three bound states, which we shall denote  as $A,B,C$, lie below the lower band, while a fourth state $D$ is located below the upper band and inside the magnetic gap. The related density profile $|\chi_\uparrow(x)|^2+|\chi_\downarrow(x)|^2$ of the bound states  is plotted in Fig.\ref{fig:Fig-Wave}(b). While the bound states $A$ and $D$  have a bell shape centered at the origin,  $B$ and $C$  exhibit two sharp lateral maxima. The eigenvalue $\lambda$ of the operator $\Lambda=\mathcal{P}\sigma_x$, shown  in Fig.\ref{fig:Fig-Wave}(a),   emphasizes  the selection rule imposed by such discrete symmetry: when the quench is performed only pairs of bound states with the same $\lambda$ can get coupled. The energy scales associated to the two transitions $A \leftrightarrow C$ and $B \leftrightarrow D$ identify two time-periods: the shorter one $\tau_{BD}=2\pi\hbar/(\varepsilon_D-\varepsilon_B) \approx 7.8 {\rm ps}$ is the period of the oscillations of the solid curve of Fig.~\ref{fig:rho}, while the smaller bound state energy difference determines the longer  period $\tau_{AC}=2\pi\hbar/(\varepsilon_C-\varepsilon_A)\approx 40{\rm ps}$ associated with an envelope of the oscillations and is hardly visible in  the density behaviour shown in Fig.~\ref{fig:rho}.\\

\begin{figure}
	\centering
	\includegraphics[width=0.95\linewidth]{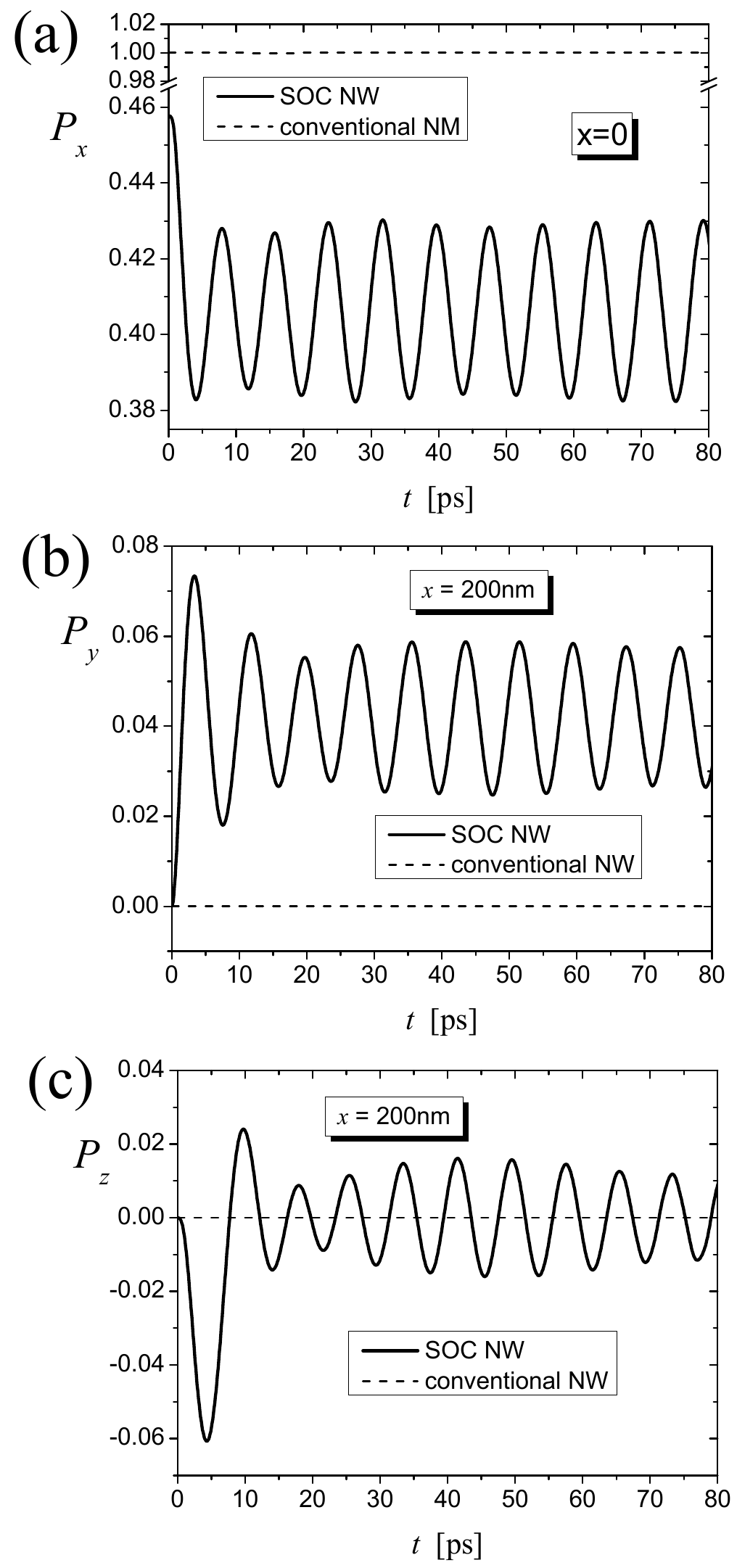}
	\caption{Time evolution of the spin polarization components after a sudden quench   of the attractive Gaussian potential (parameter values given in Sec.\ref{sub:param}): (a) the $P_x$ component at the origin $x=0$; (b)-(c) the $P_y$ and $P_z$ components at $x=+ 200\,{\rm nm}$. Solid curves refer to the case of a SOC nanowire, whose polarization components exhibit coherent oscillations. For comparison, dashed curves refer to the case of a conventional nanowire:   only  the component $P_x$ along the direction of the applied magnetic field is non-vanishing (and is constant in time), whereas the two components $P_y$ and $P_z$ orthogonal to the applied field are vanishing. }
	\label{fig:polar}
\end{figure}

{\it Spin polarization.}  Let us now turn to the spin polarization, defined as 
\begin{equation}\label{P-def}
\mathbf{P}(x) \doteq \frac{\langle \Psi^\dagger \boldsymbol\sigma \Psi^{} \rangle}{\langle \Psi^\dagger  \Psi^{} \rangle}= \,\frac{\mathbf{s}(x)}{\rho(x)}  \hspace{1.5cm} |\mathbf{P}|\le1 \quad.
\end{equation}
The time evolution of  $P_x$ (the component along the external magnetic field), shown  at   $x=0$ in the solid curve of Fig.~\ref{fig:polar}(a), also exhibits oscillations, again characterized by a period $\tau_{BD}\approx 7.8 \,{\rm ps}$.  For comparison, the dashed curve in  Fig.~\ref{fig:polar}(a) shows the behavior of $P_x$ for a conventional NW exposed to the same external magnetic field and quench potential: in that case the polarization is locked to 1 at any time and does not exhibit any oscillation. This is because  for these parameter values the spin-$\downarrow$ band is empty, and   in a conventional NW the quench potential cannot cause any coupling between the two spin sectors. 

\begin{widetext}

\begin{figure}
	\centering
	\includegraphics[width=1\linewidth]{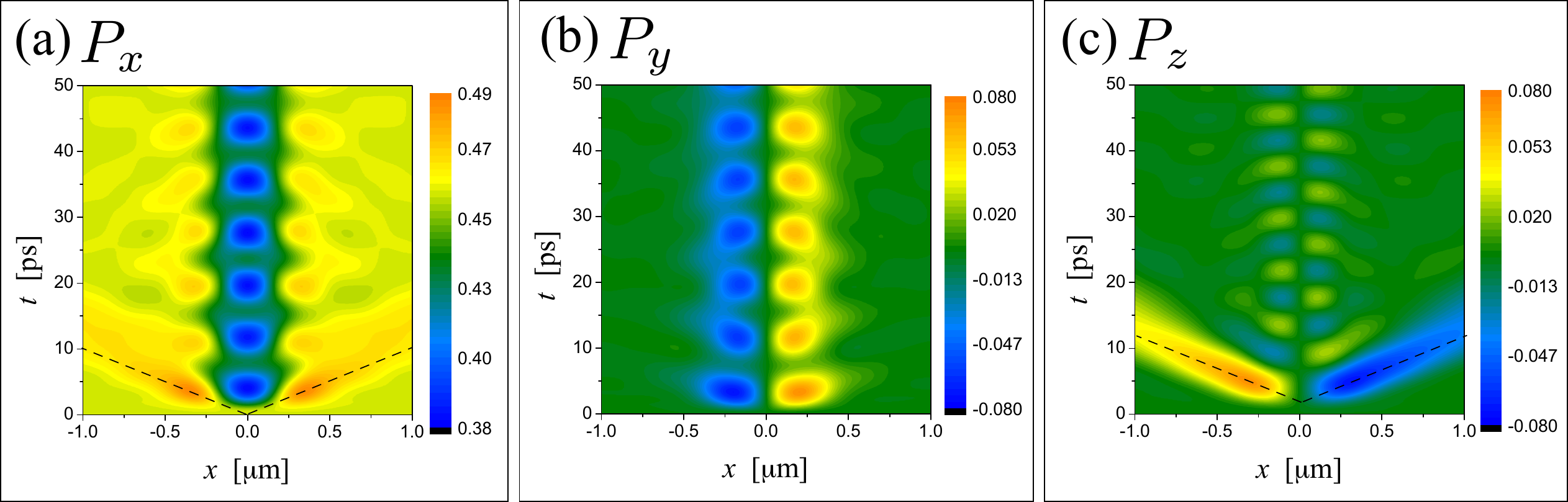}
	\caption{(Color online) Space-time colormap plots of  the three components of the spin polarization (\ref{P-def}) after a sudden quench of the impurity. Parameters are given in Sec.~\ref{sub:param}.}
	\label{fig:Fig-3D-Polarization}
\end{figure}
\end{widetext}

Concerning the  components $P_y$ and $P_z$ of the spin polarization, they  turn to be odd functions of the $x$-coordinate, so that they exist away from the origin, as shown in panels (b) and (c) of Fig.\ref{fig:polar} at a position $x=200\,{\rm nm}$. Besides the coherent oscillations with a short period $\tau_{BD}\approx 7.8\, {\rm ps}$, also  the longer envelope period   $\tau_{AC} \approx 40\,{\rm ps}$ becomes visible. Importantly, the existence of   non-vanishing spin polarization  components $P_y$ and $P_z$  that are {\it orthogonal} to the externally applied magnetic field is again a hallmark of the SOC, since in conventional NWs these components are strictly vanishing, regardless of the value of the initial chemical potential $\mu$. To provide a thorough overview of the space and time dynamics, we have plotted in Fig.\ref{fig:Fig-3D-Polarization}  a colormap of  the three components of the spin polarization (\ref{P-def}). The ``light-cone" phenomenon departing from the origin can be observed   in the $P_x$ component (dashed lines are meant as a guide to the eye), which  is spatially even around the origin and  periodically alternates between a density dip around $x=0$ and two lateral peaks located symmetrically with respect to the origin. In contrast, the $P_y$ and $P_z$ components are spatially odd around the origin. Notice that $P_y$ mainly varies in the vicinity of the applied quench potential,  with  bumps of opposite signs on the two sides of the origin, a clear hallmark of the lateral peaks of the localized wavefunctions B and C (see Fig.\ref{fig:Fig-Wave}). Moreover, also the $P_z$ component  clearly displays the  ``light-cone". In general, as compared to the electron density, the polarization exhibits more interesting features. This is because the electron density  oscillations are strongly determined by the selection rule forbidding coupling between bound states with different eigenvalue of $\lambda$, while the intrinsic spinorial nature characterizing the polarization allows coupling between different $\lambda$'s as well. 

\subsection{Interpretation of the results}
\label{sec:interpretation}
When the attractive potential is quenched, bound states can appear below both the  minima of the two NW bands, opening up the possibility for the coherent oscillation effect to emerge. However, as shown in Fig.\ref{fig:rho}, in SOC NWs the effect  appears already for much smaller values of the attractive potential strength $U_0$ as compared to the conventional NWs. In order to illustrate why this is the case, it is first  worth  showing  the following property: the interplay between the external magnetic field and the SOC effectively gives rise to an  {\em inhomogeneous} magnetic field varying over the (half of the) spin-orbit lengthscale (\ref{lSO-def}). Indeed by applying the following spin-dependent gauge transformation on the electron field spinor
\begin{equation}
\Psi(x)=\exp\left\{i\theta_{SO}(x)\sigma_z\right\}\tilde{\Psi}(x)\,,
\end{equation}
where $\theta_{SO}(x)=2x\, \mbox{sgn}(\alpha) /l_{SO}$, the original Hamiltonian is rewritten as $\mathcal{H}=\int \Psi^\dagger H\Psi^{} = \int \tilde{\Psi}^\dagger \tilde{H}\tilde{\Psi}^{}$, where
\begin{eqnarray}
\tilde{H} &=& \left(\frac{p_x^2}{2m^*}-\mu^\prime+f(t) U(x)\right)\sigma_0-\mathbf{h}_{eff}(x)\cdot\boldsymbol{\sigma}_\perp \,\,. \,\,\label{eq:effham}
\end{eqnarray}
Here $\mu'=\mu+E_{SO}$ denotes   an effective chemical potential  increased by the spin-orbit energy and $\boldsymbol{\sigma}_\perp=(\sigma_x,\sigma_y)$. Thus, in the new gauge, the Hamiltonian $\tilde{H}$  describes electrons with a customary parabolic spectrum, exposed to an effective inhomogeneous magnetic field  
\begin{equation}
\label{heff}
\mathbf{h}_{eff}(x)=h_x(\cos\theta_{SO}(x),\sin\theta_{SO}(x)) \quad,
\end{equation} 
whose magnitude $h_x$ is determined by the actual externally applied magnetic field, and whose direction rotates along the NW with a wavelength $l_{SO}/2$ determined by the SOC.  Importantly, such wavelength is tunable because the SOC can be varied e.g. by applying a backgate between NW and substrate.

\begin{figure}
	\centering
	\includegraphics[width=1\linewidth]{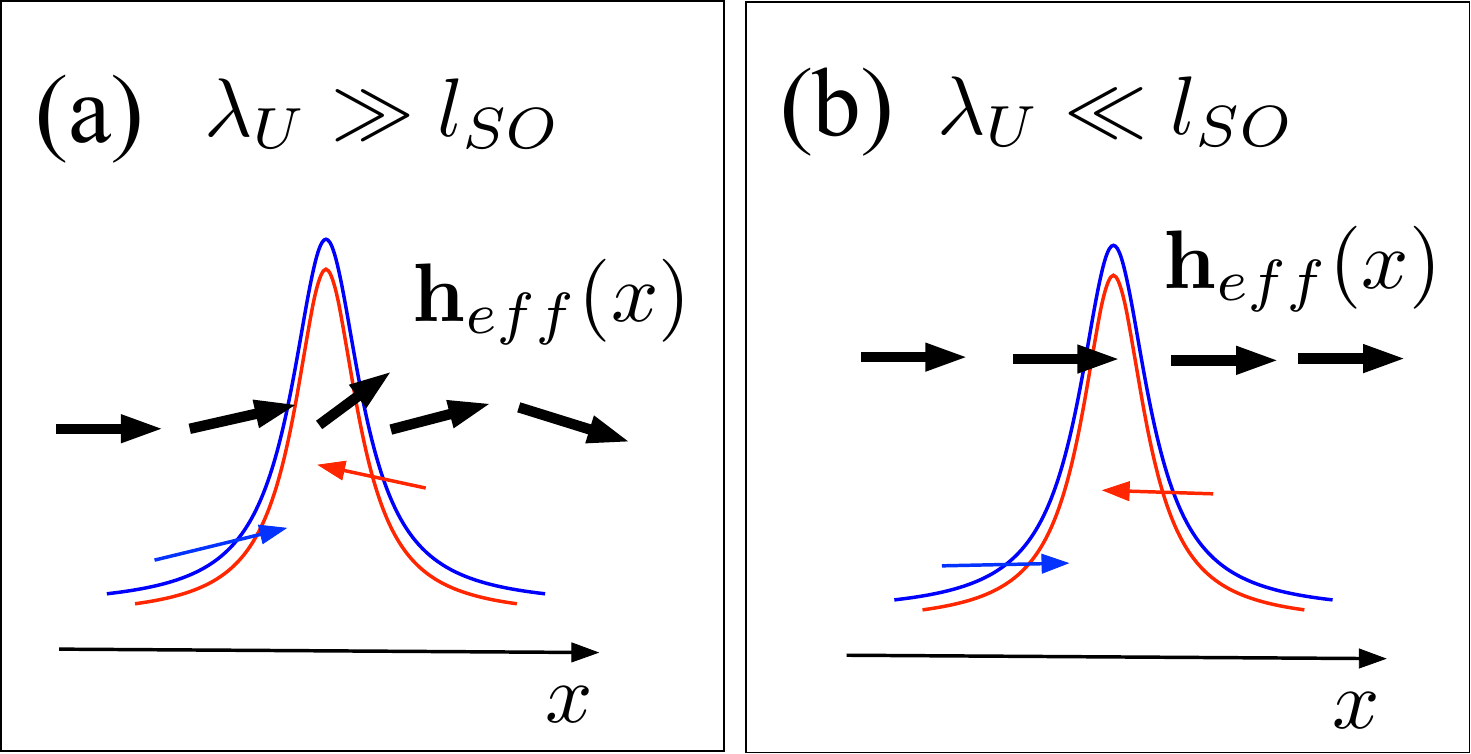}
	\caption{(Color online) The interplay between the SOC and the external magnetic field   gives rise to an effective inhomogeneous magnetic field $\mathbf{h}_{eff}(x)$ [see Eq.(\ref{heff})], whose direction rotates over the spin-orbit length $l_{SO}$, whereas the attractive quench potential extends over a lengthscale $\lambda_U$ around the origin, which also corresponds to the typical lengthscale of the bound states. (a) In the regime $\lambda_U> l_{SO}$ electron spin in the bound states experience the inhomogeneities of $\mathbf{h}_{eff}(x)$ and rapidly varies along the bound states, causing a finite overlap between different bound states and thereby leading the coherent oscillations to appear after the quench.  (b) in the opposite regime  $\lambda_U< l_{SO}$ the electron spin cannot experience the long-wavelength inhomogeneity of the effective field, so that bound states are spin-orthogonal: the spin selection causes the decay of the density.  In particular, the limit of $l_{SO}\rightarrow \infty$ corresponds to the case of conventional NWs.}
	\label{fig:fig3}
\end{figure}

Because a bound state  typically extends over the lengthscale $\lambda_U$ of the attractive potential, if the condition $\lambda_U>l_{SO}$ is fulfilled, the electron spin in the bound state  effective `sees' a spatially varying magnetic field rather than the uniform  external magnetic field, as   schematically depicted in Fig.~\ref{fig:fig3}(a). As a consequence, bound states that are spin-orthogonal in the absence of SOC, get a  finite   overlap in the presence of SOC, which causes the appearance of the coherent oscillations shown in the solid curve of Fig.\ref{fig:rho}.
In contrast, when $\lambda_U\ll l_{SO}$, the electron spin inside the bound states cannot `see'  the long wavelength inhomogeneity of the effective magnetic field, and it is essentially determined by the external magnetic field only, as sketched in Fig.~\ref{fig:fig3}(b). Their coupling tends  to vanish. Moreover,  a short-range potential causes a large momentum transfer, which displaces spectral weight from the bound state in the upper band to the continuum  states almost degenerate to it, so that the bound states in the lower band get coupled mainly to a continuum, thereby  inducing dephasing as discussed in Sec.~\ref{sec:simplemodel}.  Indeed the case of a  conventional NW   $\alpha \rightarrow 0$  in fact corresponds to the limit $l_{SO}\rightarrow \infty$, where the two bound states with opposite spin are strictly orthogonal and  oscillations are damped in the long time limit [see dashed curve in Fig.\ref{fig:rho}]. Furthermore, the inhomogeneous magnetic field $\mathbf{h}_{eff}$ also favors the formation of bound states, since the rotation of its direction along the NW causes a locking of the electron spin orientation to the orbital momentum. Thus, an electron attempting to escape the attractive potential must experience a change in its momentum, which in turn amounts to a spin mismatch, similarly to what happens in magnetic `domain walls'. The result is the enhancement of the electron localization. {Finally, the rotation of the magnetic field $\mathbf{h}_{eff}$ along the NW also explains why, despite the quench potential (\ref{Hprime}) directly acts on the charge channel, a response indirectly appears also in the spin channel and, in particular, also in polarization components that are orthogonal to the actual external applied magnetic field.}

In summary, our results can be interpreted in terms of the effective inhomogeneous magnetic field (\ref{heff}), caused by the interplay between SOC and actual magnetic field, which entails  various effects: i) it couples  the bound states related to the two bands; ii) for a fixed value of $U_0$, the number of bound states increases with the SOC; iii) it induces  coherent oscillations also in spin polarization orthogonal to the actual magnetic field.  \\

\subsection{Effects of the chemical potential}\label{sec:chem-pot}
\begin{figure}
	\centering
	\includegraphics[width=1\linewidth]{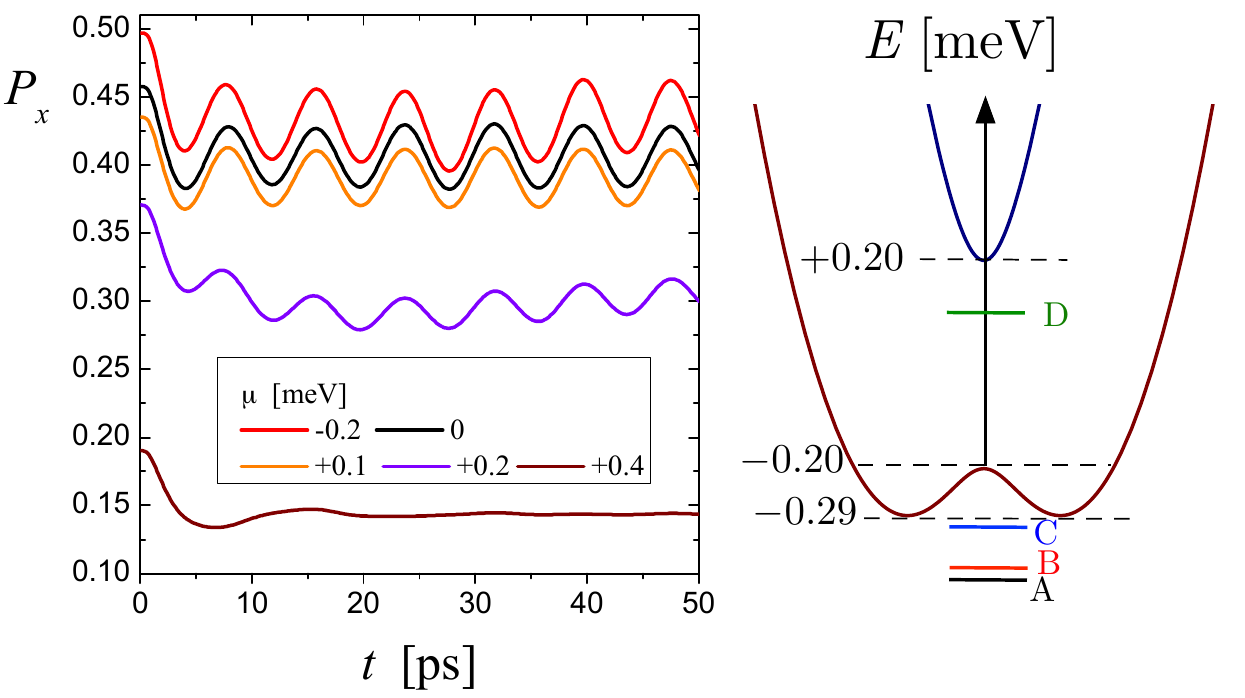}
	\caption{(Color online) Time evolution of the spin polarization component $P_{x}(t)$ (along the external applied magnetic field) after the quench, for different values of the chemical potential $\mu$ (all other parameters are   as in Sec.~\ref{sub:param}). To help locate the various chemical potential values with respect to the pre-quench energy band extrema and to the post-quench bound state energies -- shown in Fig.~\ref{fig:Fig-Wave}(a) -- they have also been reported here on the right.}
	\label{Fig-Px-muVAR}
\end{figure}

The results about the coherent oscillation effect shown in the previous sections have been obtained for chemical potential $\mu=0$, which corresponds to the center of the gap created by the external magnetic field.  Let us now address how  variations of $\mu$ affect the  phenomenon. For definiteness, we shall focus on the spin polarization $P_x$ along the NW axis, i.e. the component in the external magnetic field direction, whose time evolution is shown in Fig.\ref{Fig-Px-muVAR} for different values of the chemical potential. On the right, the position of the chemical potential values can be located with respect to the pre-quench NW band structure and the quench potential bound states.
 
As $\mu$ is increased from the center of the magnetic gap towards the bottom of the upper band, oscillations tend to be reduced. Then, for values above the magnetic gap, $\mu>h_x=0.2\,{\rm meV}$,  they are strongly suppressed. Indeed in this situation the states of the upper band are already occupied before the quench, the overlapping of the bound state below this band with continuum states becomes larger, and this ``leakage" of spectral weight ultimately results into a dephasing induced by the continuum. In contrast, when $\mu<0$, the effect of coherent oscillations  turns out to be stable. Again, we can interpret this fact as a substantial decrease of the already very small overlapping of the bound state of the upper band with continuum states.

\subsection{Effects of a finite switching time}\label{sec:finite-tau} 
So far, we have considered the customary case of a sudden quench,  where the switching time of the attractive potential is assumed to be instantaneous. We now want to analyze the effect of a finite switching time. To this purpose, we have performed calculations adopting in Eq.(\ref{Hprime}) the quench protocol
\begin{equation}\label{f(t)-def}
f(t)=\frac{1}{2}\left[1+\mathrm{Erf}\left(\frac{4(t-\tau_{sw}/2)}{\sqrt{2}\,\tau_{sw}}\right)\right]\, ,
\end{equation}
where $\mathrm{Erf}(t)$ the standard error function, and $\tau_{sw}$ identifies the   switching time of the  ramp from $f(t<0)\approx 0$ to $f(t>\tau_{sw})\approx 1$.
\begin{figure}
	\centering
	\includegraphics[width=1\linewidth]{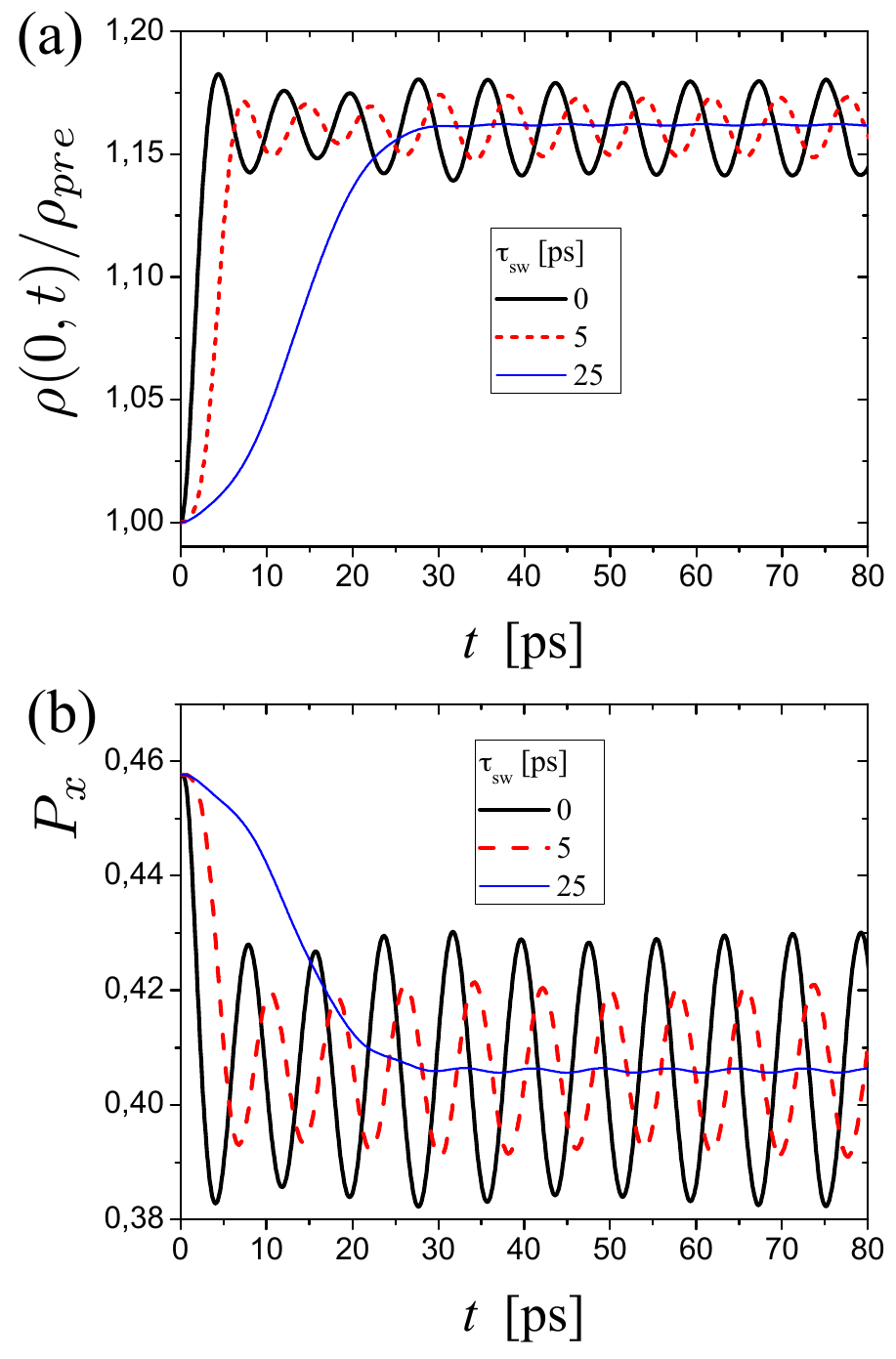}
	\caption{(Color online) (a) The time evolution of the charge density after a quench performed with different switching time $\tau_{sw}$ of the ramp protocol function (\ref{f(t)-def}). While the thick black curve corresponds to the case of sudden quench ($\tau_{sw}\rightarrow 0$) analyzed in the previous section, the dashed red curve  and the thin blue curve describe the behavior for $\tau_{sw}$ comparable and much longer than the coherent oscillations period $\approx 8\,{\rm ps}$, respectively. (b) The same as panel (a) for  the  $P_x$ component of the  spin polarization.}
	\label{fig:Fig-switch}
\end{figure}
In the case of a sudden quench, corresponding to $\tau_{sw} \rightarrow 0$, the oscillations have a period of about $8\,{\rm ps}$. In Fig.~\ref{fig:Fig-switch}(a) we compare the electron density behavior obtained for the sudden quench to the cases of two smooth ramps with $\tau_{sw}=5\,{\rm ps}$ and $\tau_{sw}=25\,{\rm ps}$. One concludes that oscillations are robust as long as the switching time is shorter than the oscillation period, and their visibility is slightly reduced (see dashed red curve).  Only when $\tau_{sw}$ is much longer than such oscillation period (thin blue curve), the oscillations are strongly damped, as shown by the thin blue curve. Fig.~\ref{fig:Fig-switch}(b) shows the same comparison for the $P_x$ component of the spin polarization, with similar results as the charge density.\\
 
\section{Conclusions}\label{sec:conclusions}
In this work we have shown that quenching a localized attractive potential in a NW can induce   coherent oscillations in the dynamics of the   observables, which thus do not relax to a steady state value. The origin of this phenomenon is twofold: on the one hand the spectrum, which is    purely a continuum of states before the quench,  undergoes a 'topological' change across the quench by acquiring a discrete portion arising from the formation of bound states. On the other hand, the pre-quench state is a complex excitation of the post-quench Hamiltonian. While the continuum part of the spectrum   would lead to a saturation of the long-time dynamics, the bound states form a sort of ``dephasing-free" subset with an internal oscillatory dynamics, and the period of these oscillations is related to the energy separation between different bound states. Selection rules impose to have at least three bound states for this   effect to occur. In conventional NWs, this requires that the quench potential must be sufficiently strong  (see Fig.\ref{fig:Fig-ESO=0}).

The conditions to observe the effect are much more favorable in SOC NWs exposed to an external magnetic field, which naturally offer a twofold advantage, as we have shown in  Secs.\ref{sec:socwire} and \ref{sec:results}. First,  the effect arises already for weak attractive potential strength (see Fig.\ref{fig:rho}). Second, the quench of the potential, which directly couples to the charge sector, also indirectly yields non-trivial coherent dynamics in the spin sector as well, and coherent oscillations can be observed in all components of the spin polarization, including the ones that are orthogonal to the applied magnetic field  (see Figs.\ref{fig:polar} and \ref{fig:Fig-3D-Polarization}).  The space-time patterns of spin polarization is particularly insightful, for it carries traces of the bound state wavefunctions.   
The emergence of these remarkable features is due to   the interplay between the external magnetic field $h_x$ and the SOC, which effectively generates a inhomogeneous magnetic field, whose magnitude depends on $h_x$ and whose direction rotates over a lengthscale determined by the spin-orbit length $l_{SO}$. In particular, in the regime where the lengthscale of the localized quench potential is larger than the spin-orbit length, $\lambda_U>l_{SO}$, the effective magnetic field favours the formation of bound states and also causes a coupling between bound states that would otherwise be decoupled  [see Sec.\ref{sec:interpretation}]. 

While the results presented here were given for the case of a Gaussian attractive potential, the method we used is straightforwardly generalizable to other confining potential. We have discussed   the effect  of the variation of the chemical potential, showing that the situation where it lies in the middle or below the magnetic gap ($-E_{SO} \lesssim \mu <0$) seem to be particularly suitable for the observation of the effect (see Fig.\ref{Fig-Px-muVAR}). Furthermore, while most results were presented for the case of a sudden quench, in Sec.\ref{sec:finite-tau} we have explicitly analyzed the effect of a finite switching time of the ramp protocol function (\ref{f(t)-def}), showing that the coherent oscillation effect is robust  as long as the switching time $\tau_{sw}$ is smaller than (or comparable to) the oscillation period (see Fig.\ref{fig:Fig-switch}).  

In terms of experimental implementations,  we observe that the local attractive potential can be realized by applying a narrow metallic gate near   the NW, whose geometrical size determines the lengthscale $\lambda_U$ of the quench potential~\cite{kouwenhoven_2012,liu_2012,heiblum_2012,xu_2012,defranceschi_2014,marcus_2016,marcus_science_2016}, while the applied gate voltage implements the potential strength $U_0$ utilized here (see Fig.\ref{fig:Fig-setup}). Importantly,  the SOC coupling can be tuned with various  techniques,~\cite{weperen,gao-2012,nygaard_2016,sasaki_2017,loss_2017} even by some order of magnitude, enabling one to directly control the spin-orbit length. These features thus provide various handles to control the formation of the bound states and the period of the coherent oscillations related to their energy separation.  Finally, we observe that, as in any quantum quench,   the predicted oscillations can be observed as long as their period is much smaller than the characteristic decoherence time for charge or spin degrees of freedom due to environment. As shown in the previous section, for typical parameters one has $\tau\approx 5-10\ \mathrm{ps}$. For InSb and InAs wires, values of $\tau_{\phi}>100\ \mathrm{ps}$ have been estimated~\cite{srel1,srel2} for the spin degrees of freedom at low temperatures. 
Moreover, recent experiments on cold atoms have shown that SOC one-dimensional systems can be also realized also using laser beams connecting the hyperfine levels  in  Fermi gases.~\cite{soc3}
The predicted effect can thus be observable   in experimentally realistic situations, and may open up interesting scenarios for the investigation of quantum quenches in spintronics.


\begin{thebibliography}{99}
\bibitem{powerlaws} It can be checked that both terms decay with a power law~$t^{-1}$.
\bibitem{valSO} {In some figure we will also present results for the case $E_{SO}=0.05\,{\mathrm meV}$ with corresponding $l_{SO}=220\,{\mathrm nm}$.}
\end{thebibliography}
\end{document}